# Build Smart Grids on Artificial Intelligence—A Real-world Example

**Shutang You[1] , Yilu Liu[1,2], Hongyu Li[1], Shengyuan Liu[1], Kaiqi Sun[1], Yinfeng Zhao[1], Huangqing Xiao[1], Jiaojiao Dong[1], Yu Su[1], Weikang Wang[1], Yi Cui[1]**

**Address: 1. University of Tennessee, Knoxville, U.S.**
**2. Oak Ridge National Laboratory, U.S.**
**Email: syou3@utk.edu**

**Abstract:** Power grid data are going "big" with the deployment of various sensors. The big data in power grids creates huge opportunities for applying artificial intelligence technologies to improve resilience and reliability. This paper introduces multiple real-world applications based on artificial intelligence to improve power grid situational awareness and resilience. These applications include event identification, inertia estimation, event location and magnitude estimation, data authentication, control, and stability assessment. These applications are operating on a real-world system called FNET/GridEye, which is a wide-area measurement network and arguably the world-largest cyber-physical system that collects power grid big data. These applications showed much better performance compared with conventional approaches and accomplished new tasks that are impossible to realized using conventional technologies. These encouraging results demonstrate that combining power grid big data and artificial intelligence can uncover and capture the non-linear correlation between power grid data and its stabilities indices and will potentially enable many advanced applications that can significantly improve power grid resilience.

---

## 1. Introduction of the AI Application Platform — FNET/GridEye

The power grid is a large-scale cyber-physical system. It's also critical infrastructure that many other industries and facilities depend upon. Due to its importance, many sensors, including the SCADA system sensors, fault recorders, smart meters, power plant sensors, and synchrophasor measurements, etc. have been deployed to the power grid to increase its situational awareness. With the increasing number of high-data-rate sensors, power grid monitoring systems provide unprecedented detailed wide-area information to assist real-time situational awareness and decision making in smart grids [1, 2].

With the increase of the number of sensors and the growing demand of high-quality situational awareness from the industry, the limitations of conventional methods to process and digest data started to emerge across multiple dimensions, such as accuracy, speed, robustness, and scalability [3, 4]. With extensive experience on the power grid, sensor development, and data analytics, the FNET/GridEye team (http://fnetpublic.utk.edu/) has been intensively exploring, developing, and deploying artificial intelligence (AI) technologies in FNET/GridEye for a decade. Many successful applications based on AI have been developed on FNET/GridEye and used by U.S. power utility companies and regulatory entities [5-7].

---

Email: Shutang You (syou3@utk.edu)

Developed using synchrophasor measurement technologies, FNET/GridEye is a wide-area power system monitoring system that covers worldwide power grids. (Its predecessor is called FNET, which is the abbreviation of Frequency Monitoring Network [8].) The sensor of FNET/GridEye is called Frequency Disturbance Recorder (FDR) [9-12], as shown in Figure 1. Plugged into outlets, FDRs take frequency, voltage magnitude and phase angle, and power quality measurements from the power grid. Additionally, FDR has a GPS module to provide accurate timestamps to its time-series measurement data [9]. The data collected by the FDR are sent through the Internet to the data center, and then processed in the data server and digested in web servers, real-time application servers, post-event analysis and storage servers, and backup servers [13]. Figure 2 shows the overall architecture of FNET/GridEye. As of 2020, it has around 300 sensor units deployed around the world (around 200 of them are in the U.S.). The distribution of FDRs in the continental U.S. and the worldwide coverage of FNET/GridEye are shown in Figure 3.

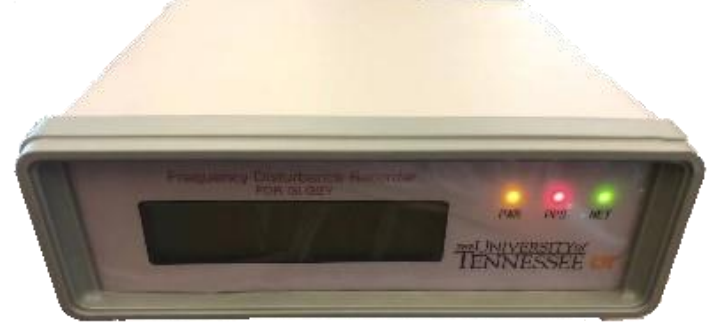

**Figure 1.** Frequency disturbance recorder (FDR).

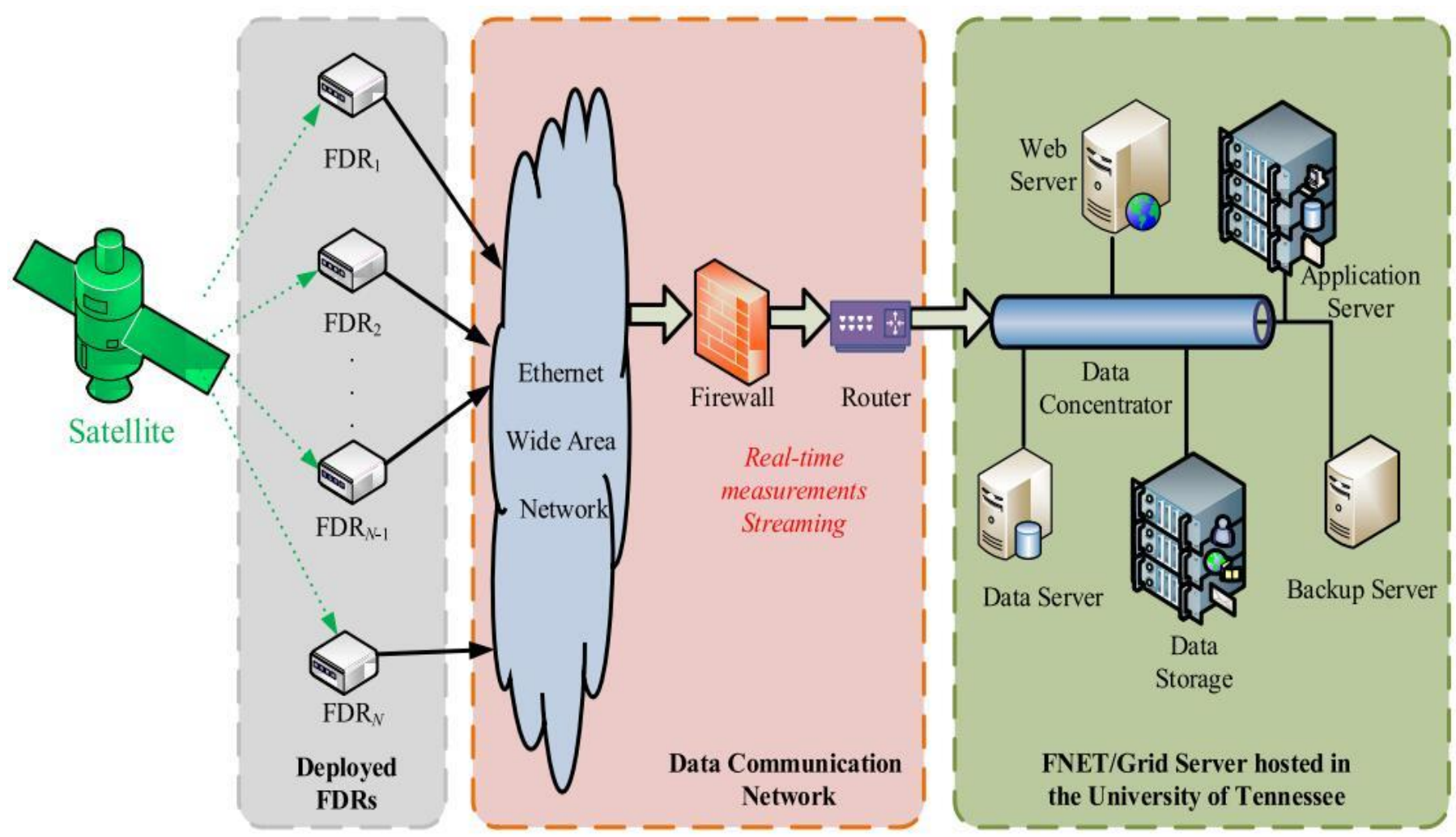

**Figure 2.** FNET/GridEye architecture.

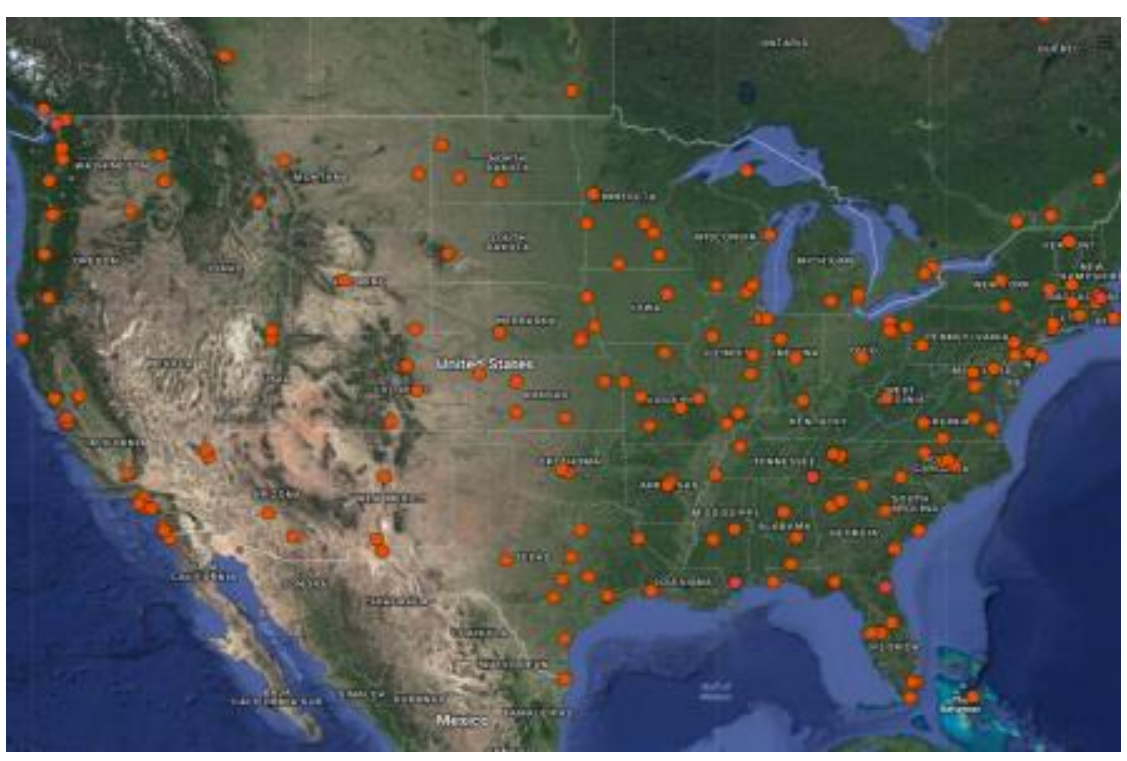

**(a)** FDRs deployed in North America.

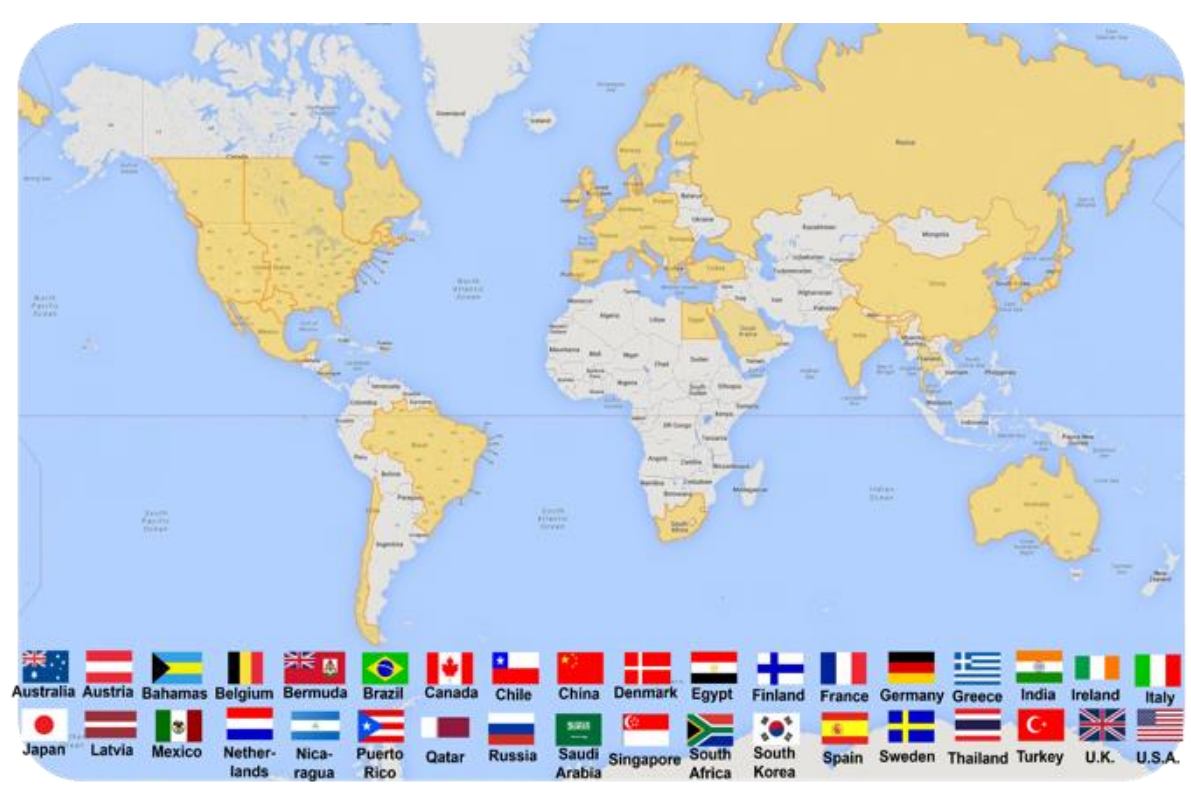

**(b)** Countries with FDRs deployed.

**Figure 3.** FDR deployment in the U.S. and worldwide.

FNET/GridEye measurement contain valuable information that can indicate the system dynamic behaviors and system health. Based on the distribution-level frequency and voltage measurement collected by FNET/GridEye, a series of applications has been developed to assist power grid situational awareness. As an example of these applications, Figure 4 shows the event report automatically generated by the FNET/GridEye system after a generation trip event followed by electromechanical oscillations. This event report visualized the frequency measurements from each location, and more importantly, provided the estimated event location and magnitude. The frequency measurement collected near the generation trip location had a large dip down to around 59.96Hz, and quickly bounced back due to the support from the rest of the grid. Meanwhile, the system frequency declined to 59.98Hz before frequency regulation facilities (Automatic Generation Controls) brought the frequency back 60Hz gradually. Other applications on FNET/GridEye include the frequency and voltage phase angle monitoring, oscillation detection, model validation, islanding detection, generation trip and load shedding detection, line trip detection, load control, data statistical analysis, and cyber-attack detection, etc. [13-27].

Basic Event Information

| Event Date | Event Time | Event Type | Estimated Amount |
| --- | --- | --- | --- |
| 2018-08-06 | 16:12:47 UTC | Generation Trip | 690 MW |
| Point A | Point B | Point C | Point C Prime |
| 60.009 Hz | 59.9868 Hz | 59.9809 Hz | N/A Hz |
| MOD-027-1 Event | Inter Connection | Estimated Reliability Coordinator | ROCOF |
| NO | EI | MRO | N/A |
| Estimated Event Location | | Additional Location Information | |
| (47.2611, -93.6528) | | near Clay Boswell power plant (MRO) in (Cohasset,MN,55721). | |
| Unit Detection Order (the first 6 units) | | | |
| UsMnGre790, UsMnMisostpaul832, UsMnDodgeCenter877, UsNdMdudickinson729, UsNdMduwilliston726, UsMTMduglendive730 | | | |

**Figure 4.** Event identification report from FNET/GridEye.

This paper will introduce several AI-based applications based on the FNET/GridEye platform. These applications include event identification, inertia estimation, event location estimation, data authentication, frequency control, and stability assessment (Figure 5). These AI-based applications are operating on FNET/GridEye servers (each with 2 Intel Xeon CPU E5-2470 processers and 128 GB memory) shown in Figure 2.

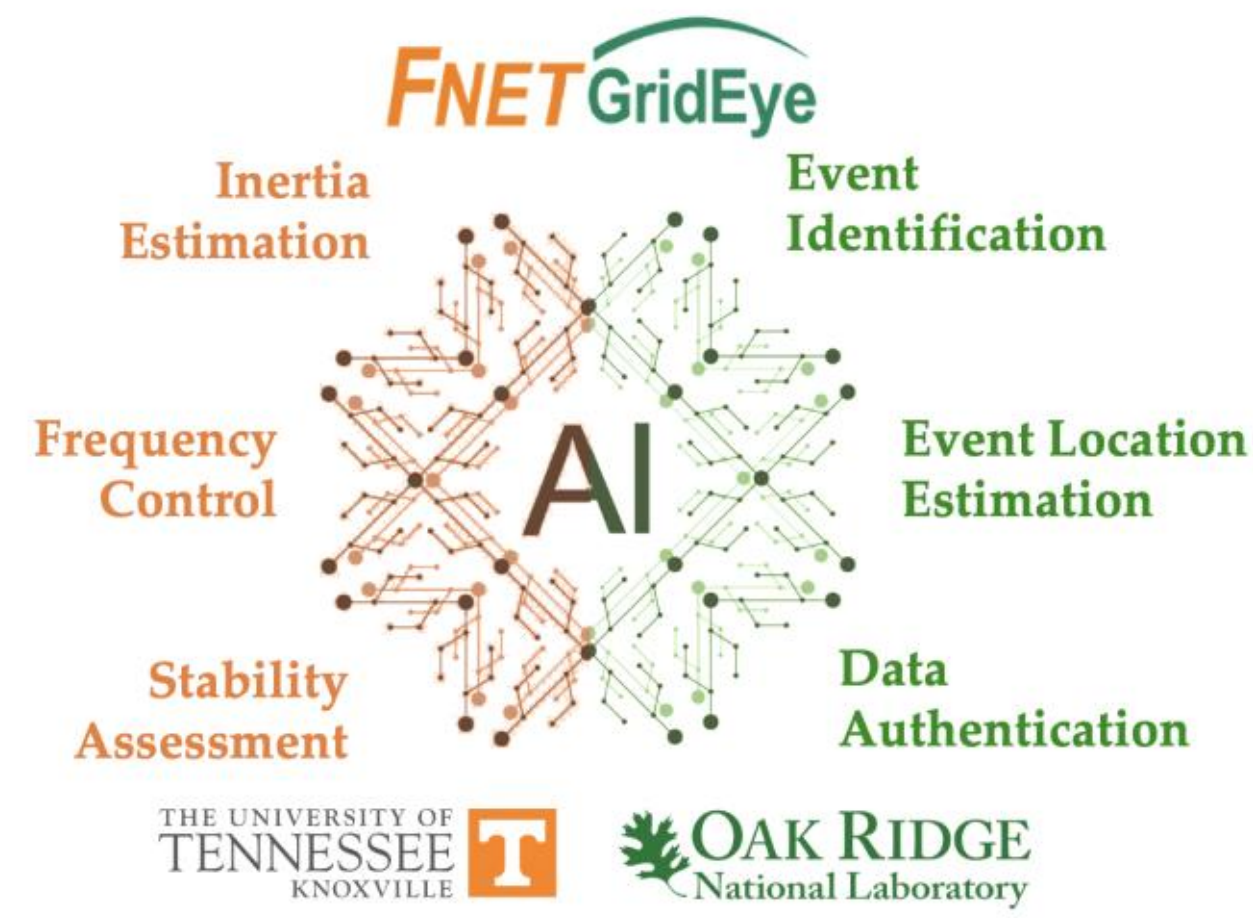

**Figure 5. AI-based applications on FNET/GridEye**

## 2. Event Identification Based on AI [28]

Power grid events, for example, generator trips, occur on a daily basis [29]. One of the FNET/GridEye real-time applications is to detect power grid events and provide stake holders real-time event reports, including detailed information on event types, locations, and magnitudes. One challenge in real-time event identification is to distinguish actual grid events from regular power grid regulation processes as they have similar footprints seen from measurements [28]. For example, a generation trip is the disconnection of one or more generators due to a malfunction that triggers the protection relay; whereas frequency ramping up/down is caused by the regulation of generator governors to respond to system load changes. Figure 6 shows a comparison on the frequency and the unwrapped and aligned voltage phase angle for a generation trip event and a frequency ramping down event. It can be seen that they have similar patterns in both frequency and voltage phase angle footprints. In the generation trip event, the real and reactive power of the generator reduces to zero instantly after the trip. For the frequency ramping event, the system imbalance happens in a less abrupt way. Both events result in a decline in the system frequency within a short period.

To improve the accuracy of grid event identification and mitigate the impact of interference from non-event incidents, an AI-based approach, which consists of two convolutional neural network models and classifier fusion, is used to detect the event in FNET/GridEye. The measurement data from FNET/GridEye, including frequency and angle data, are first constructed into an image, as shown in Figure 7. (Both colored and gray images can be utilized. The colored image codes the image by converting the input data into RGB values, whereas the gray image converts input data into gray grades.) Then, two convolutional neural networks (CNN) that separately intake the images constructed by rate-of-change-of-frequency (RoCoF) data and relative angle shift (RAS) data as inputs are developed to detect the event independently. Then, these two CNN classifiers are fused to synthesize their advantages in detecting different types of events to give the likelihood degree of whether an event belongs to a specific event type, which is the output of AI. (More details on constructing the CNN model can be found in [28].) The accuracy and time consumption for the proposed and conventional methods in event identification are shown in Table 1. It is seen that although AI spends a little more time on detecting each event, it is much more accurate than the conventional method, which is based on setting thresholds for the frequency deviation magnitude and the rate of change of frequency.

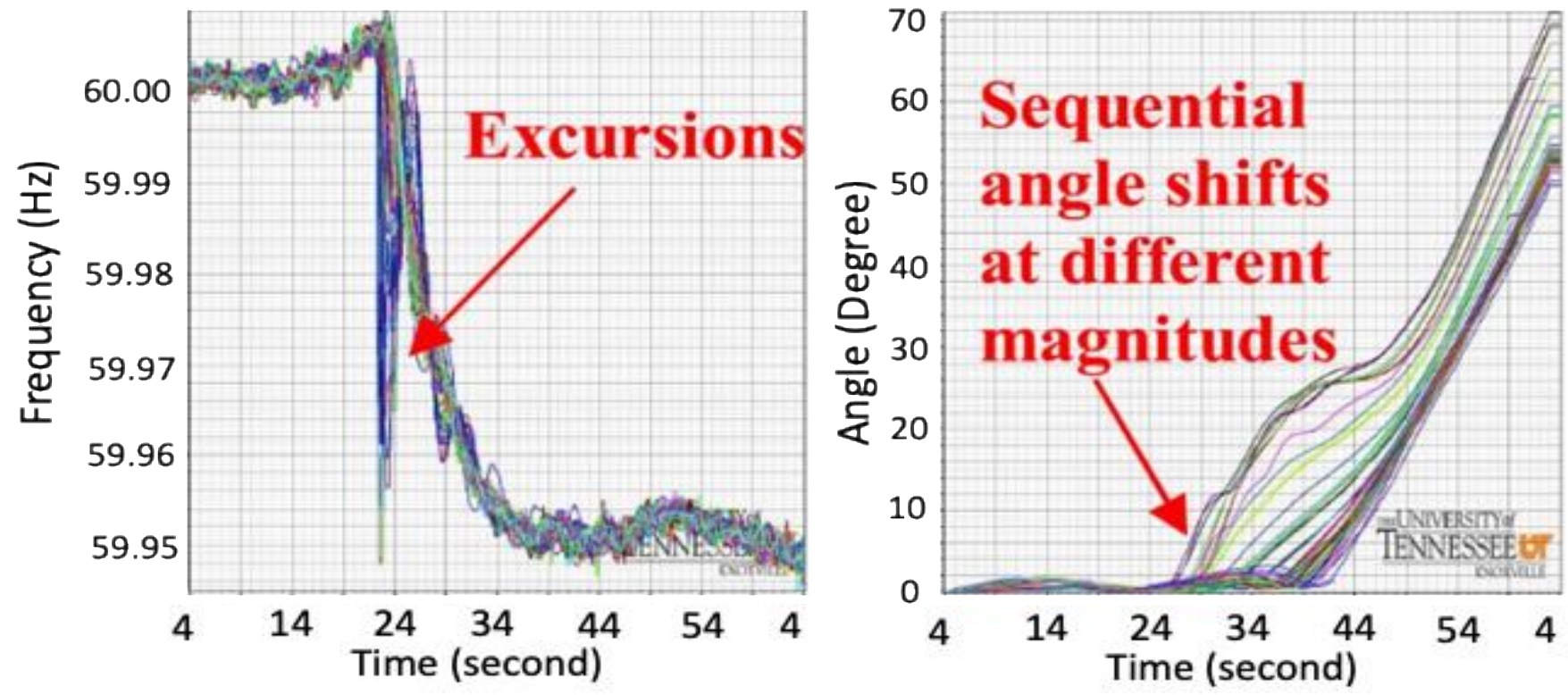

(**a**) A generation trip event (left side: frequency; right side: unwrapped relative angle)

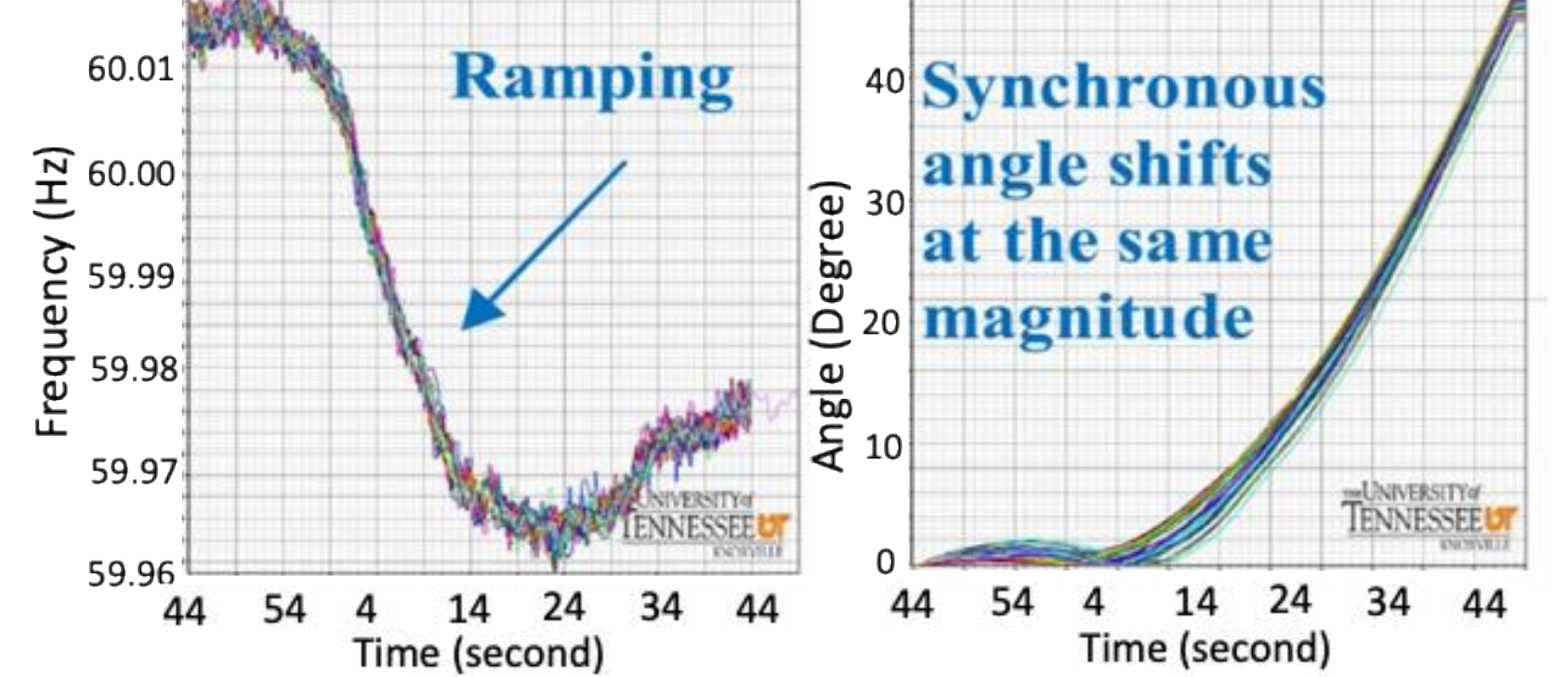

(**b**) Frequency ramping down (left side: frequency; right side: unwrapped relative angle)

**Figure 6.** Frequency and unwrapped voltage phase angle footprints of (a) a generation trip event and (b) a frequency ramping down event

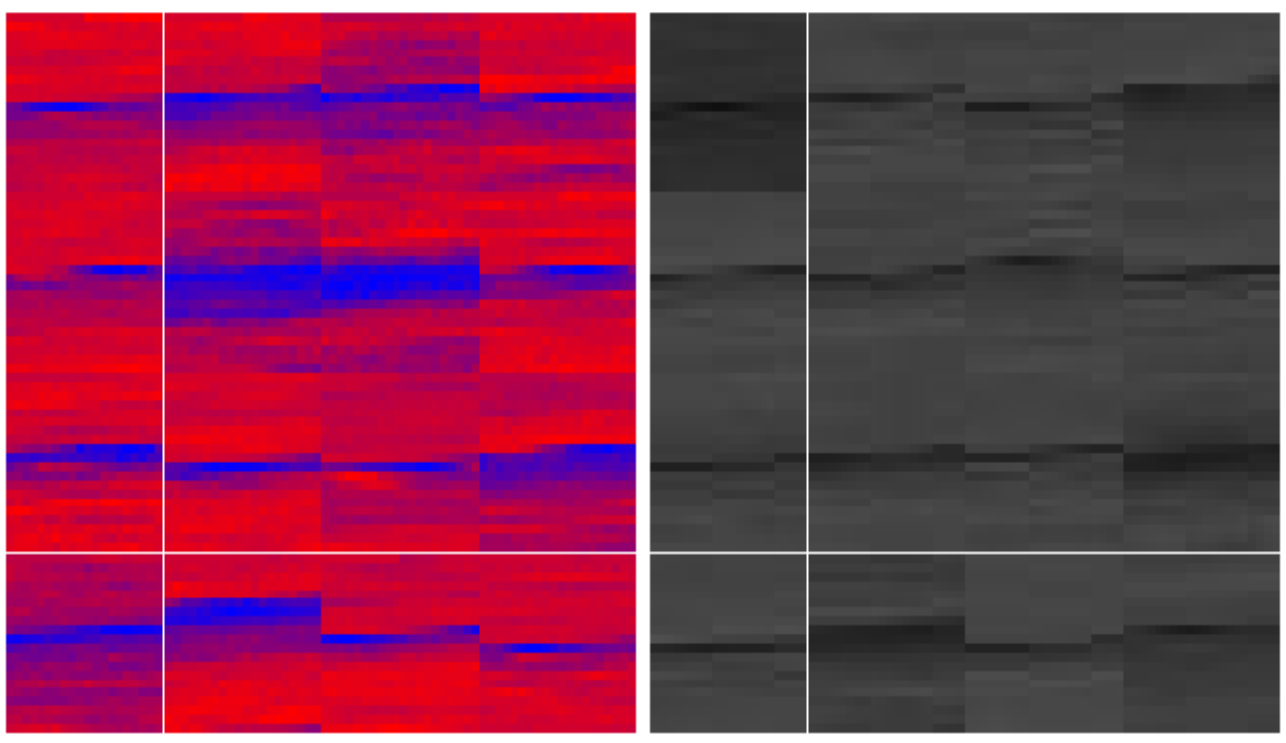

**Figure 7.** Color image (left) and gray image (right) converted using generation trip event data from FNET/GridEye.

**Table 1.** Accuracy and time comparison for event identification.

| | Conventional method | Proposed Model (Color image based) | Proposed Model (Gray image based) |
|---|---|---|---|
| Identification Accuracy | 51.1% | 97.8% | 97.8% |
| Identification Time (ms) | 0.010 | 0.810 | 0.812 |

## 3. AI-based Inertia Estimation Using Ambient Synchrophasor Measurement [30]

Power system inertia is the kinetic energy stored in mechanical rotors of synchronous generators and motors. It is a critical system parameter that determines the power grid stability after major disturbances such as generation loss and load shedding. A power system with large inertia is more robust to disturbances, while a small power system or a high renewable system will have large frequency fluctuations after the same disturbances due to small inertia. This, in turn, will further result in under-frequency-load-shedding and generation trip. With the increase of renewable generation, inertia in most power grids is decreasing continuously and is commonly seen as an increasingly important attribute to monitor [5, 14, 31-34].

To monitor system inertia in real-time, an AI-based inertia estimation method is developed using FNET/GridEye frequency measurement data. This application takes advantage of the ambient variations in the frequency measurement obtained from multiple locations. The change of the system inertia will influence the relative magnitudes and phases of these oscillations, which can be quantified by the minimum volume enclose ellipsoid (MVEE) method. Then, the quantitative metrics of the MVEE can be utilized as input features of the AI model. An example of the system ambient oscillations and MVEEs at different inertia levels are shown in Figure 8. The system inertia in Figure 8 (a) is two times of the system inertia in Figure 8 (b). It can be seen that MVEEs (on the right side of Figure 8) can extract the differences in the system states from the ambient frequency variations. These differences can be further utilized as inputs by random forests to estimate the system inertia, which is the AI output. The predicted and actual inertia of the WECC system are shown in Figure 9. (The actual inertia is obtained by adding the inertia of all on-line units together.) It can be seen that AI can obtain accurate inertia estimation for the power system. Since the inertia estimation method uses ambient oscillation information, it is more accurate and effective in large-scale power systems.

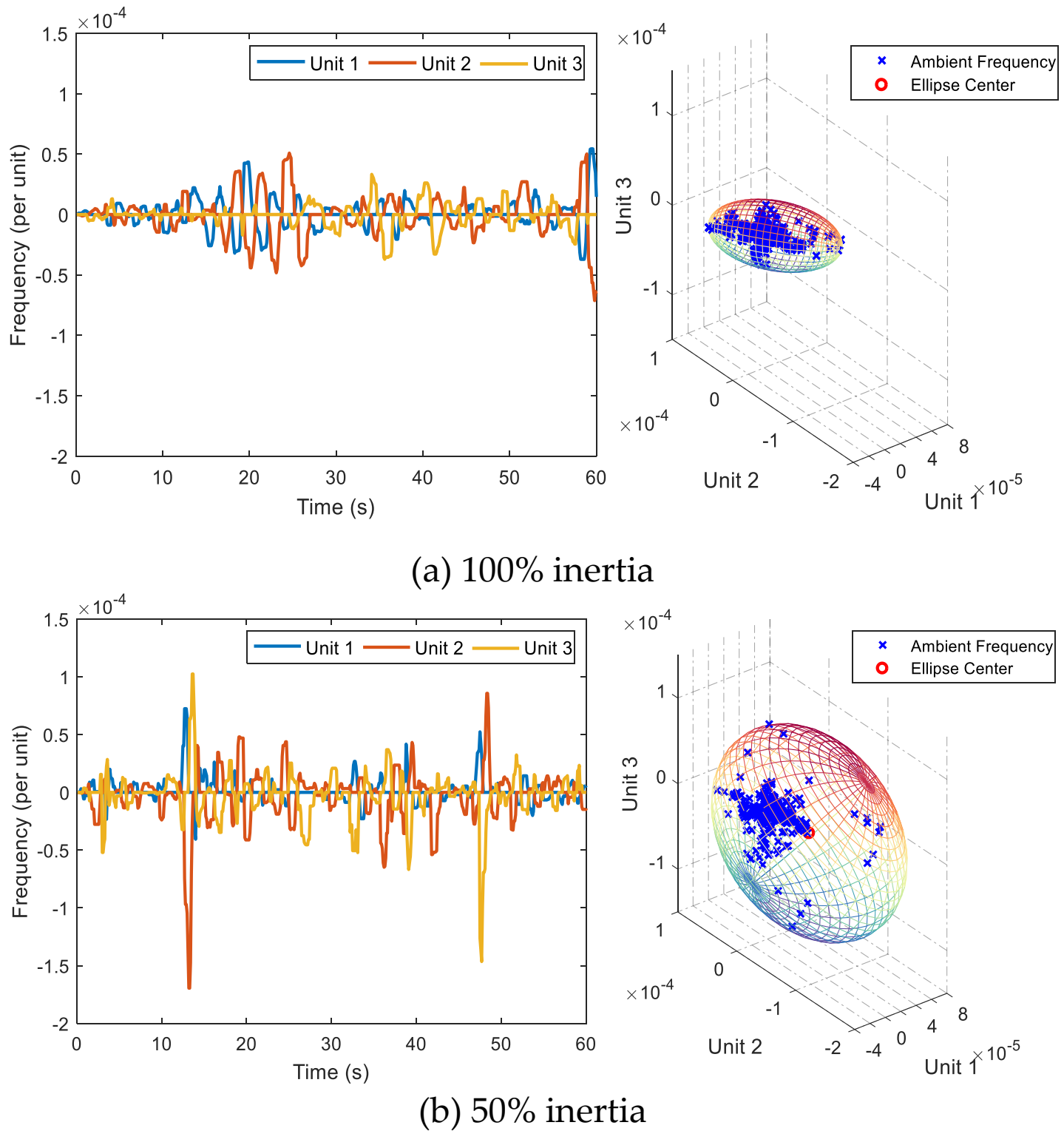

**Figure 8.** Ambient frequency variation (left) and characteristic ellipsoids (right) at (a) 100% and (b) 50% inertia level.

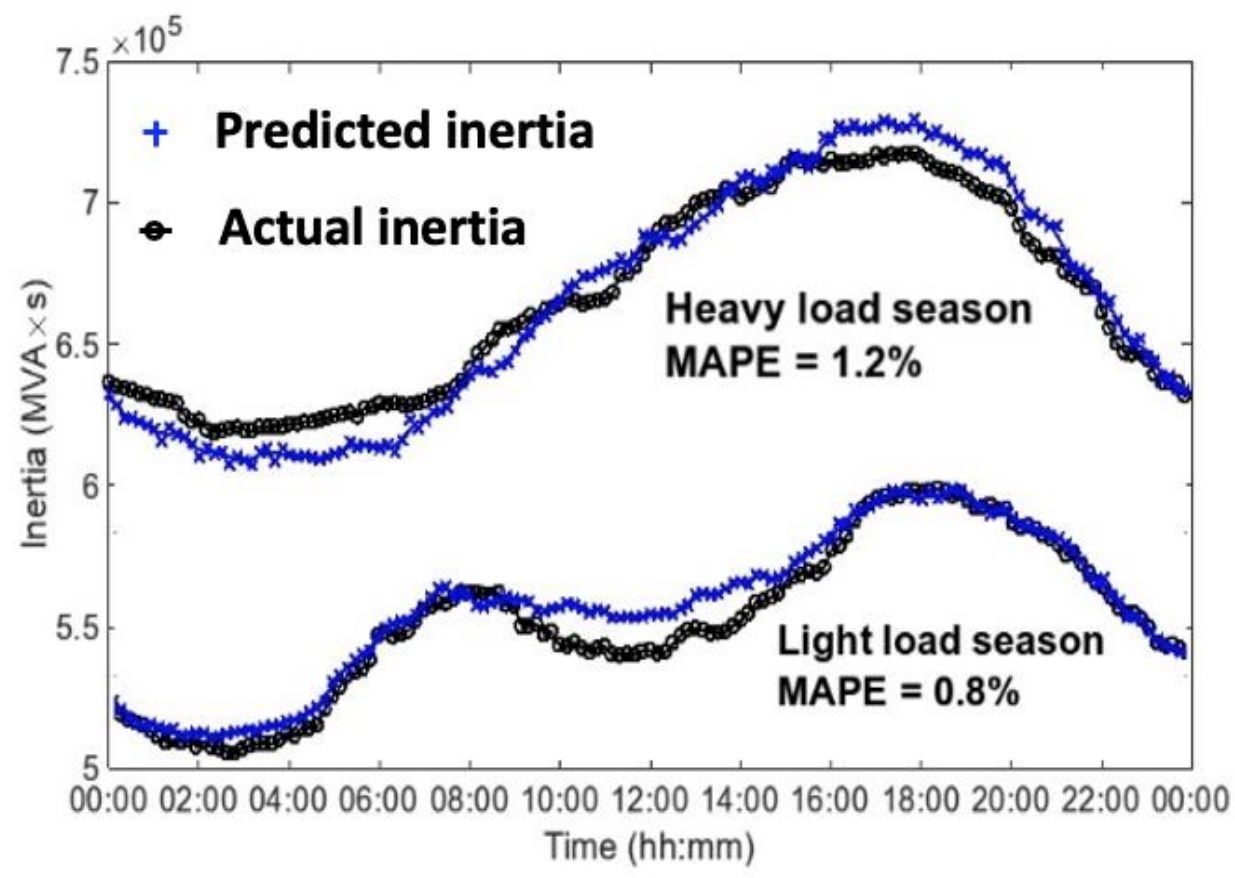

**Figure 9.** Comparison of measured and estimated inertia in the WECC system during heavy and light load seasons.

## 4. AI-based Event Location and Magnitude Estimation

The location and magnitude of power grid events are important information for system operators [35-39]. AI technologies can also be utilized to estimate event location and magnitude. Figure 10 shows MVEEs constructed for four time segments of event frequency measurement data from FNET/GridEye during a generation trip event. W1 is the measurement at the beginning of the event. W2 is the initial period of the event. W3 and W4 are the intermediate and the settling periods of the event respectively. MVEEs are applied to frequency measurements to extract quantitative information from multiple measurements at different stages of an event. Then, the extracted information from MVEE is used as inputs in random forests to learn the relation between the quantitative information and event location/magnitude. The outputs of random forests are event location and magnitude. Figure 11 shows the accuracy comparison of the AI-based event location method and the conventional method based on Time-Delay-of-Arrival (TDOA) [18]. Table 2 and Table 3 show the disturbance location and magnitude estimation accuracy compared with conventional methods. It can be seen that the AI-based method has higher accuracy in both event location estimation and magnitude estimation.

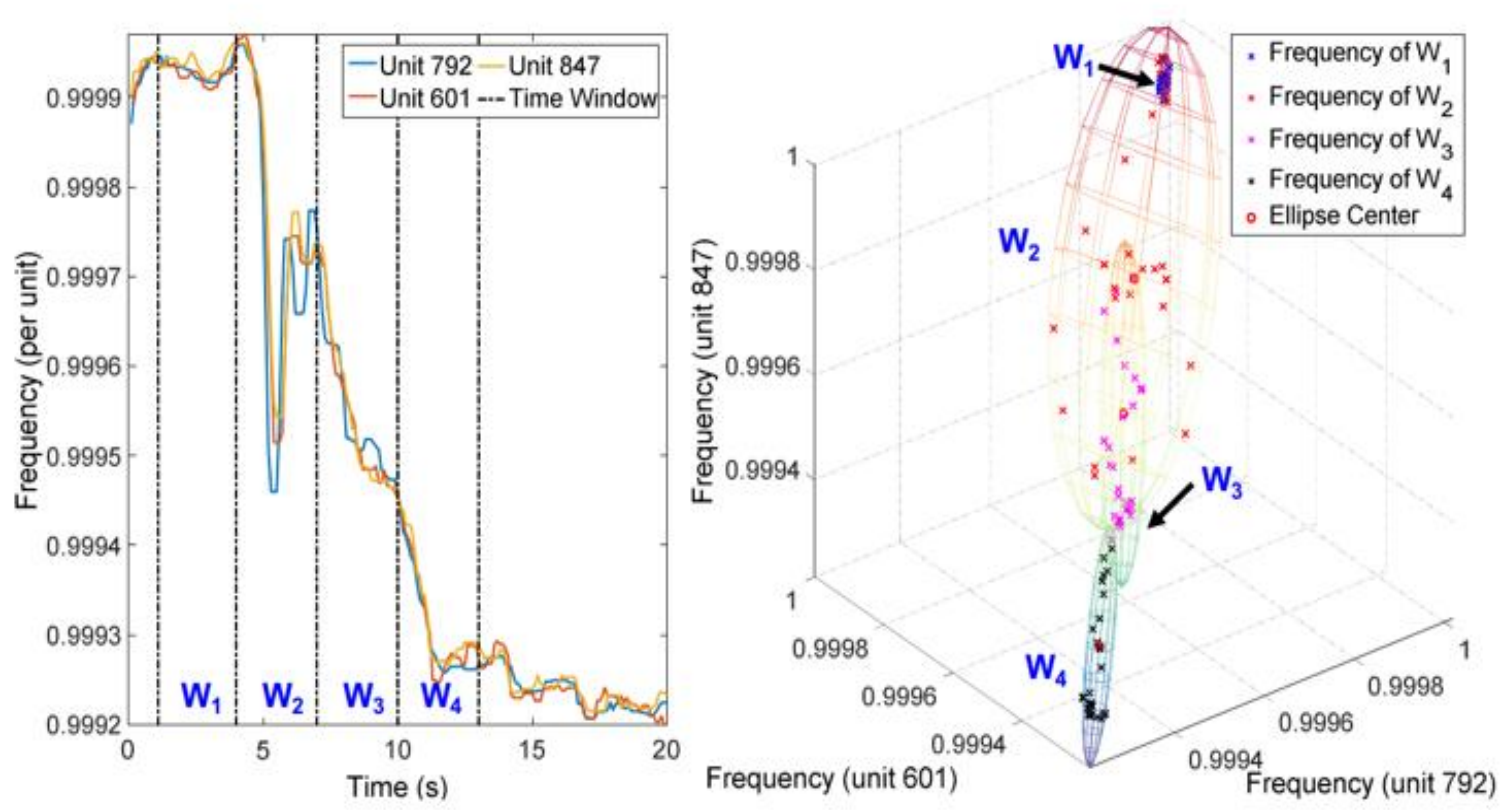

**Figure 10.** Frequency ellipsoids during the generation trip starting at 4 second.

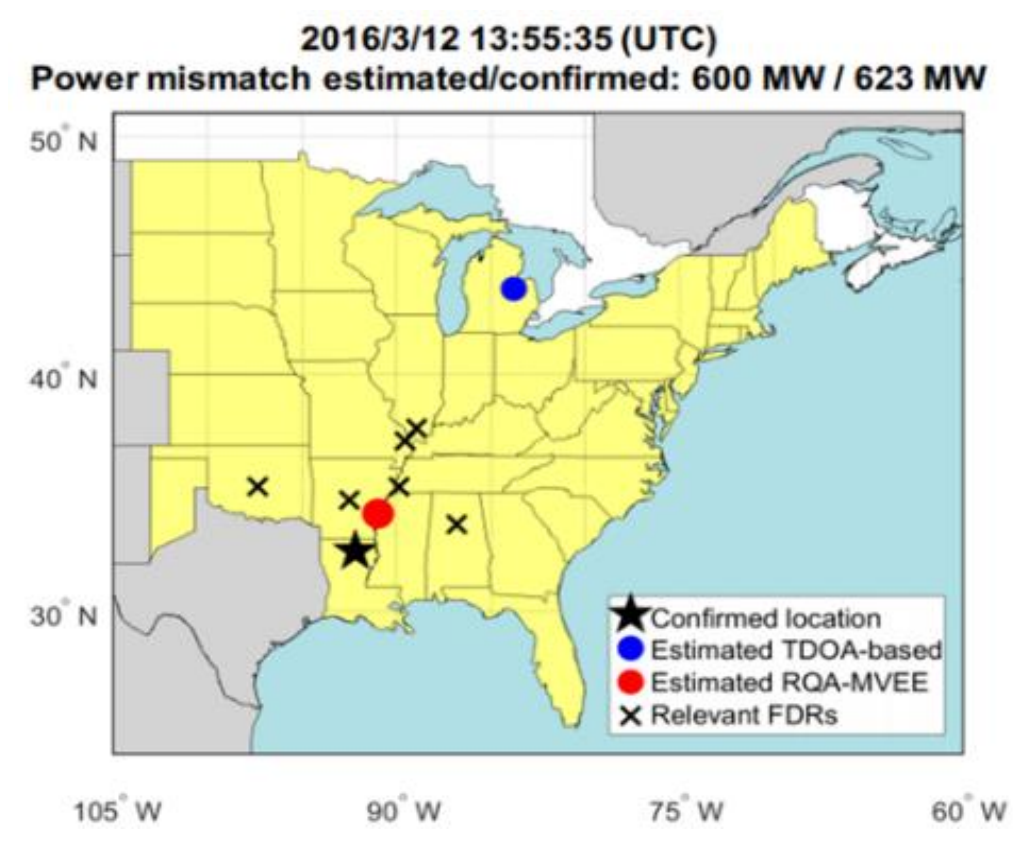

(**a**) Event 1 on March 12, 2016

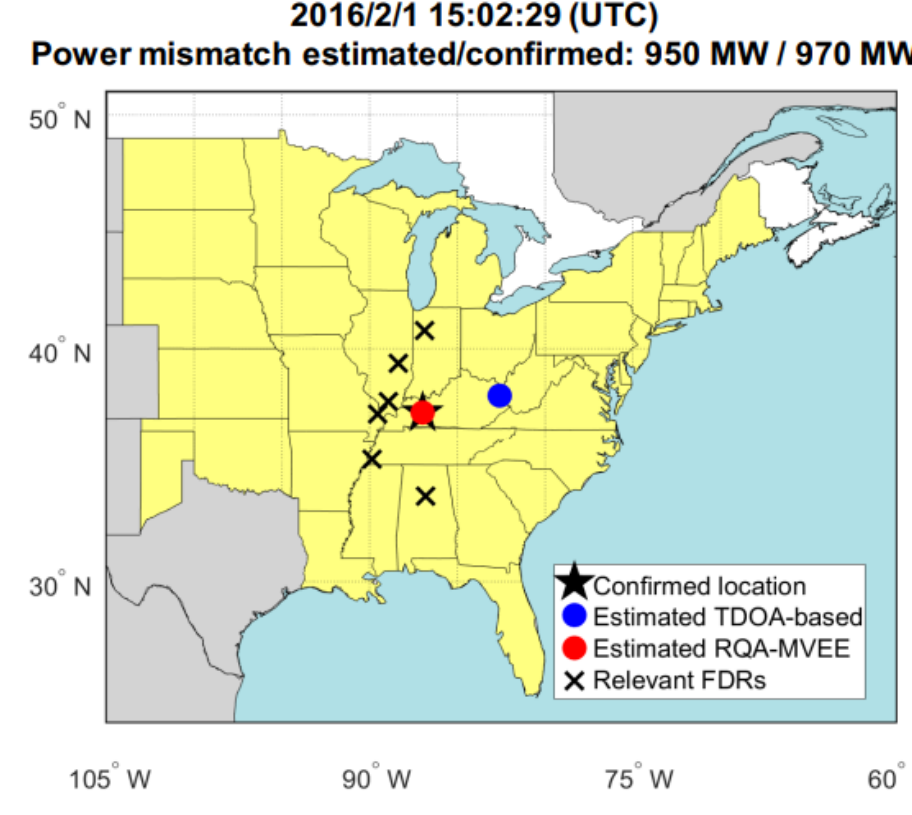

(**b**) Event 2 on February 1, 2016

**Figure 11.** Comparison of disturbance locations estimated by TDOA-based and AI method.

**Table 2.** Disturbance location estimation comparison.

| **Location estimation error (miles)** | **TDOA-based method** | **AI-based method** |
|---|---|---|
| | **(percentage of events)** | |
| 0 | 30% | 70% |
| <50 | 50% | 98% |
| <100 | 65% | 100% |

**Table 3.** Power mismatch estimation comparison.

| **Mismatch estimation error (%)** | **Beta value-based method** | **AI-based method** |
|---|---|---|
| | **(percentage of events)** | |
| <10 | 45% | 80% |
| <20 | 70% | 95% |
| <30 | 95% | 100% |

## 5. Model-Free Data Authentication Using AI [40]

The importance the cybersecurity increases with the wide applications of wide-area measurement systems in system situational awareness and control. As data are collected from sensors installed at grid edges and transmitted through the wide-area network, these data are vulnerable to various cyber-attacks [18, 41, 42]. This is also one of the reasons that the conventional communication infrastructure for remote monitoring and control is physically isolated from other civil communication networks. With the development of internet-of-things, more sensors and controllers are integrated into the public communication network such as the Internet. This physical isolation is being broken and the security of the power grid measurement system is under an increasing risk of cyber-attacks.

To mitigate this risk, AI can be used to authenticate data collected from wide areas to ensure the data are not compromised and spoofed. The advantage of AI is that it can comprehensively identify multiple characteristics in measurements to authenticate the data source. The input pattern selection is based on the physical principles and testing on different input features. In this AI-based data authentication application, the measurement data components at different time scales are extracted using a time-frequency signal processing technology named ensemble empirical mode decomposition (EEMD) [40]. Then, the multiple extracted components at different frequency ranges are analyzed using Fast Fourier Transform (FFT). The frequency spectrum of FFT is used as input of the BP neural network as a classifier (Figure 12). The output is the match degree of the measurement with a specific FDR ID.

A test was done using three closely-located FDRs. Figure 13 shows the locations of three FDRs deployed in Knoxville, Tennessee, U.S. The average distance of these three FDRs is 7.9 km. The frequency measurements from these FDRs with a 1.44 kHz data reporting rate are used for data authentication. A cyber-attack scenario is emulated by swapping the frequency measurement of these FDRs. Since all frequency measurements are real-time data and appear identical at adjacent locations, it is challenging to identify this data-spoofing attack. The proposed AI method achieved 80.9% accuracy in authenticating the data source, significantly higher than other methods, whose accuracy is around 60-70%.

It is noted that in this application, the number of hidden layers can be adjusted based on the number of input data. A sensitivity study on the number of hidden layers shows that 2-hidden layer BP neural network has better performance than 1-hidden layer neural network (77% accuracy) in this case study. Further increasing the number of hidden layers will add little improvement but significantly increase the computational burden.

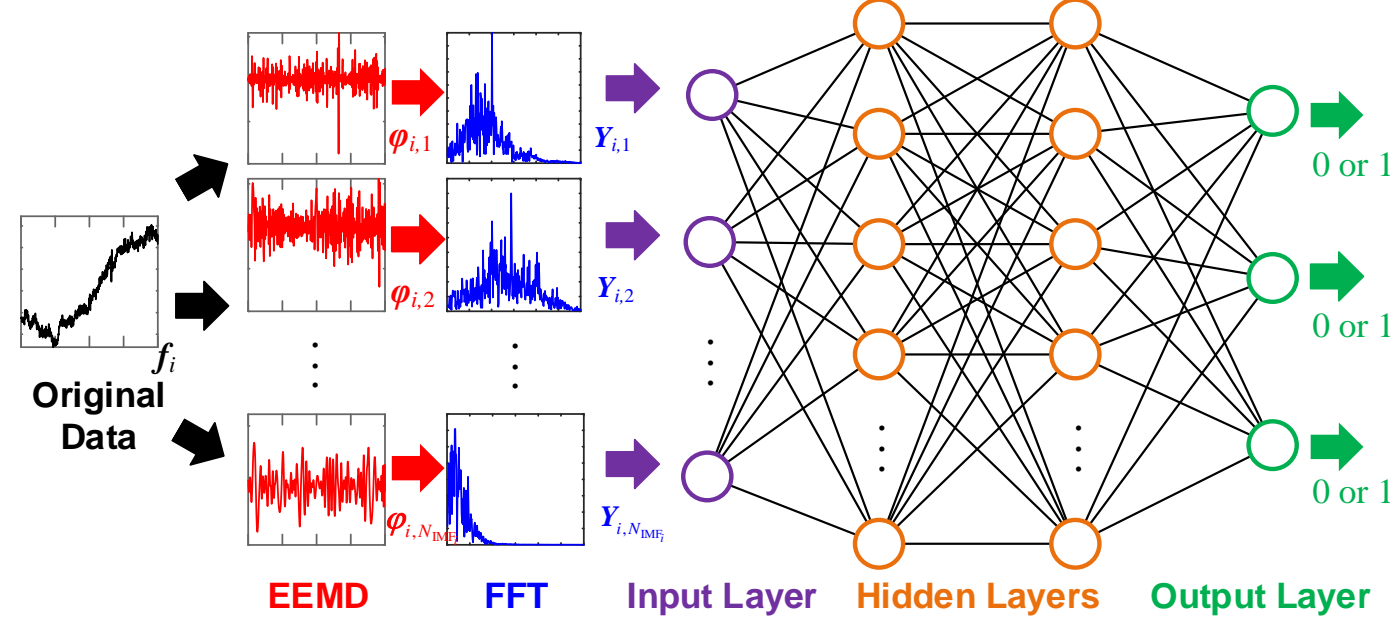

**Figure 12.** Architecture and input features of the proposed BP neural network.

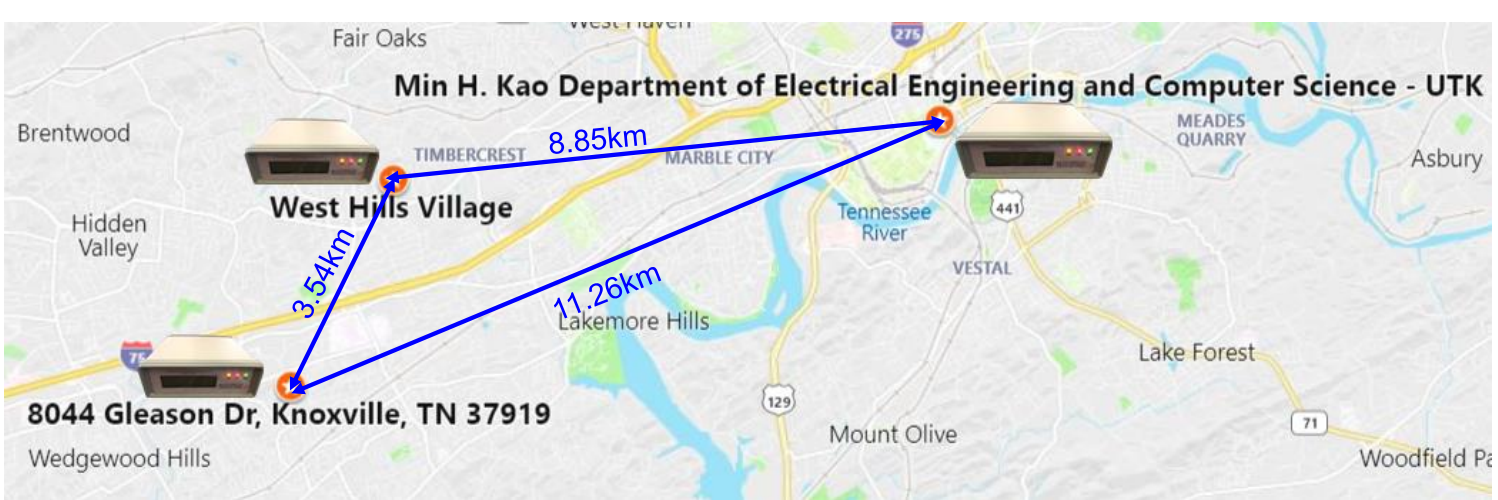

**Figure 13.** Locations of three FDRs.

## 6. AI-based Frequency Control [43, 44]

AI can not only help situational awareness, but also further improve power grid controls [45, 46]. With the increase of renewable generation and HVDC facilities, the frequency control of the power grid increasingly relies on more advanced technologies that can better utilize system information and prescribe corresponding control strategies as appropriate. Figure 14 shows the flowchart of an advanced PV farm frequency control using AI. In this frequency control, AI is used in two places: one is estimating system real-time inertia; the other is estimating the PV headroom requirement to ensure the system has enough power reserve to maintain frequency stability after the largest generation loss. The estimation of inertia using AI has been introduced in Section 3. The method to estimate the PV headroom reserve requirement is shown in Figure 15. In this method, the neural network is used to develop a model to predict the frequency control performance for a specific set of system parameters, which include system inertia, system governor response, and PV headroom reserve. The output of AI is the system frequency nadir. Then, the trained AI model is combined with a binary-search method to iteratively find the minimum reserve requirement for a specific control target. The performance of the AI-based frequency control method is shown in Figure 16. It demonstrates that the system frequency can be maintained above the 59.8 Hz threshold to avoid under-frequency-load-shedding (UFLS), regardless of the changes in system conditions, including PV output change shown by the yellow line in Figure 17. The difference between the green bars and blue bars in Figure 17 represents the savings of PV power reserve for the AI-based frequency control. It shows that the proposed AI-based frequency control can save a substantial amount of solar energy.

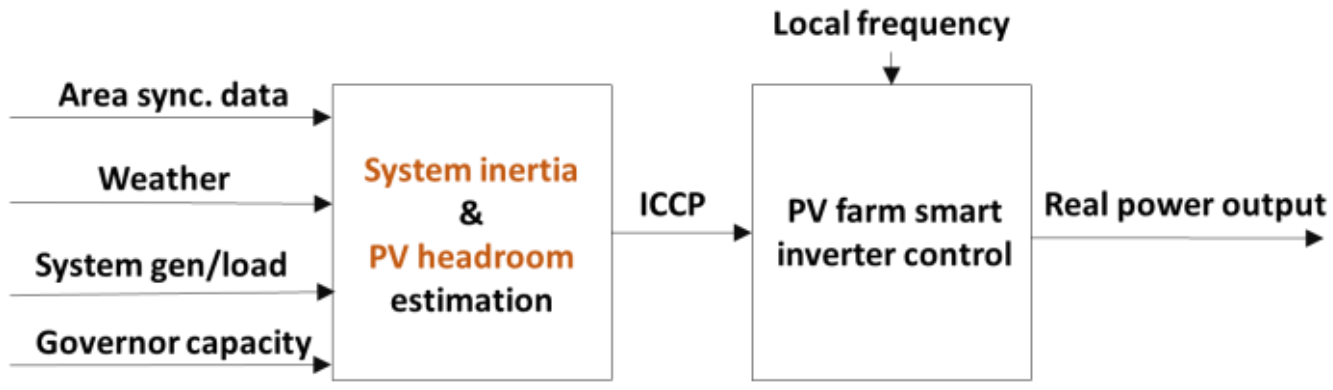

**Figure 14.** AI-based PV frequency control.

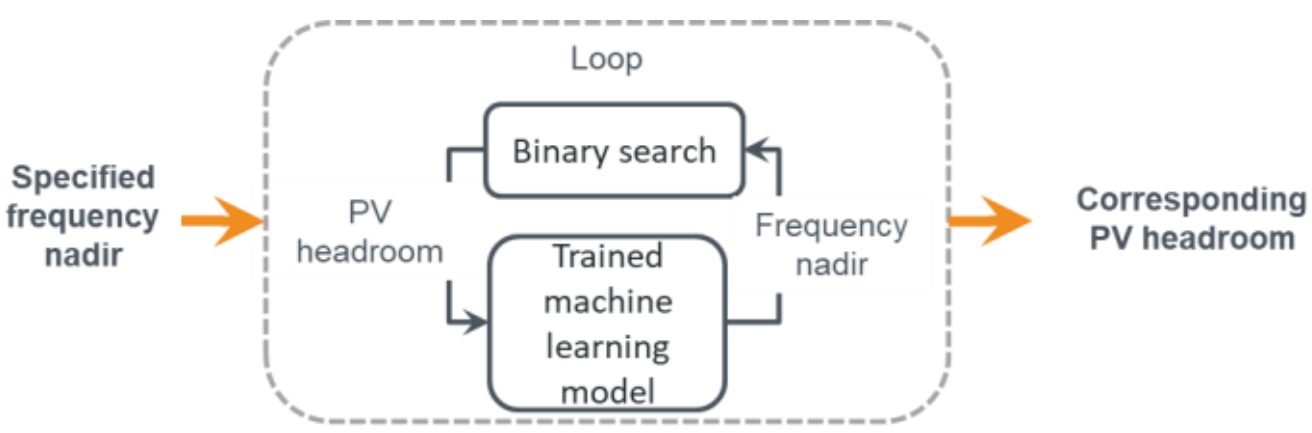

**Figure 15.** AI-based PV headroom requirement estimation.

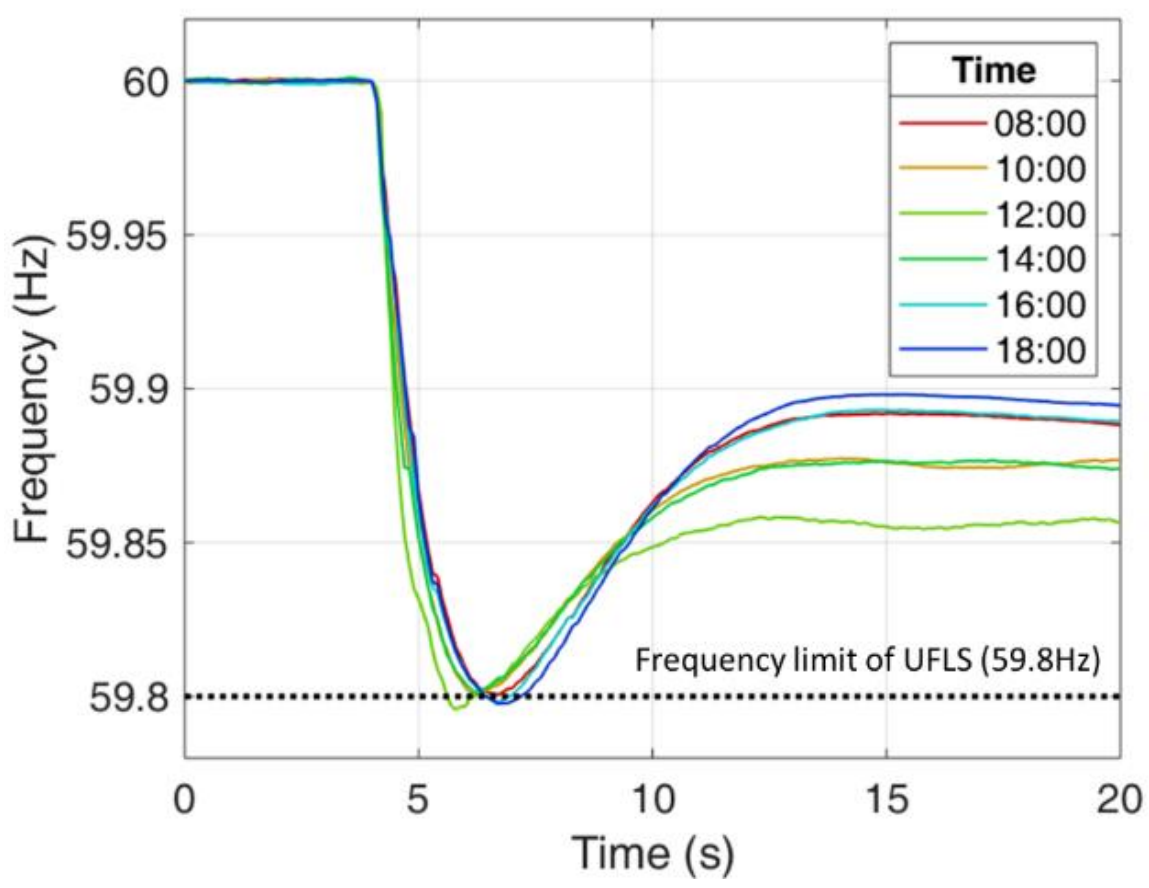

**Figure 16.** AI-based PV frequency control performance for different hourly scenarios.

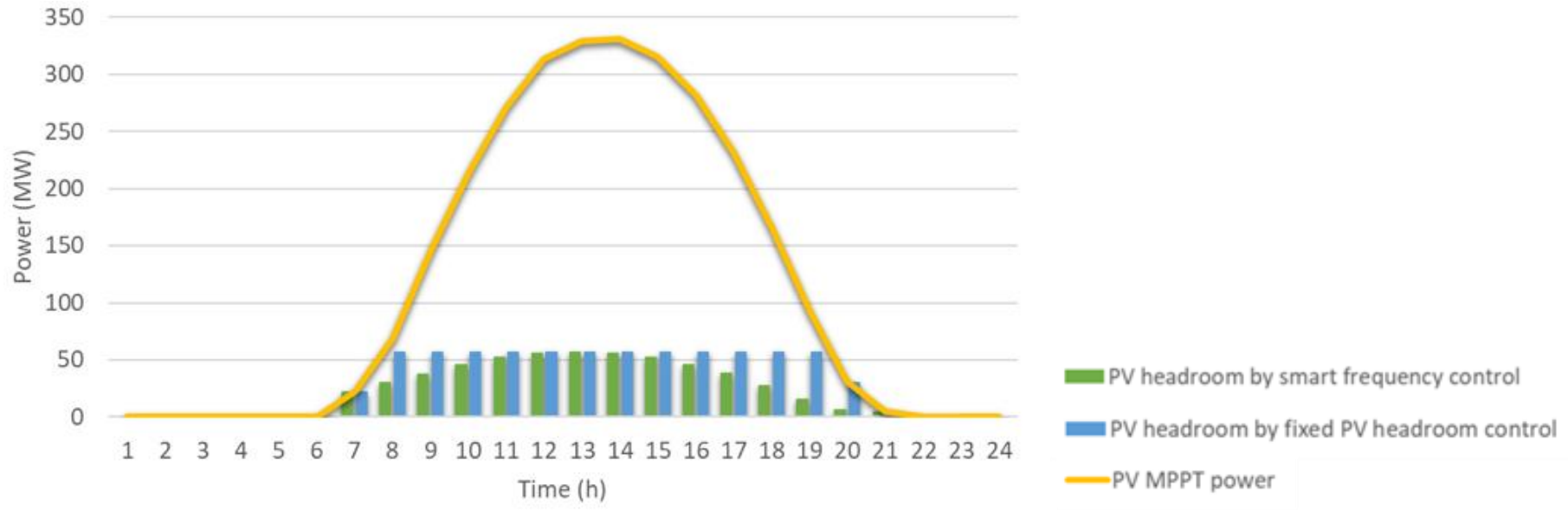

**Figure 17.** PV headroom (curtailment) of AI-based frequency control and fixed PV headroom control.

## 7. Stability Prediction Based on AI [47]

Besides frequency stability, power system stability also includes transient stability, small signal stability, and voltage stability [48]. These stabilities are important metrics of the systems' capability to maintain dynamic security after disturbances, such as short circuit faults, line tripping, and oscillations [49]. Quantifying the stability margin is helpful for operators and planners, because it can assist decision making to steer around these unstable and risky states to reduce system risk. The challenge of quantifying these stability margin is the availability of the dynamic models and the computation time for real-time stability assessment [50-52].

**Table 4.** Input and output of stability assessment.

| Stability | Input | Output |
|---|---|---|
| Frequency | Generation dispatch results, inertia | Frequency nadir |
| Transient | Generation dispatch results, transmission network | Critical clearing time (CCT) |
| Small-Signal | Generation dispatch results, transmission network | Damping ratio + frequency |

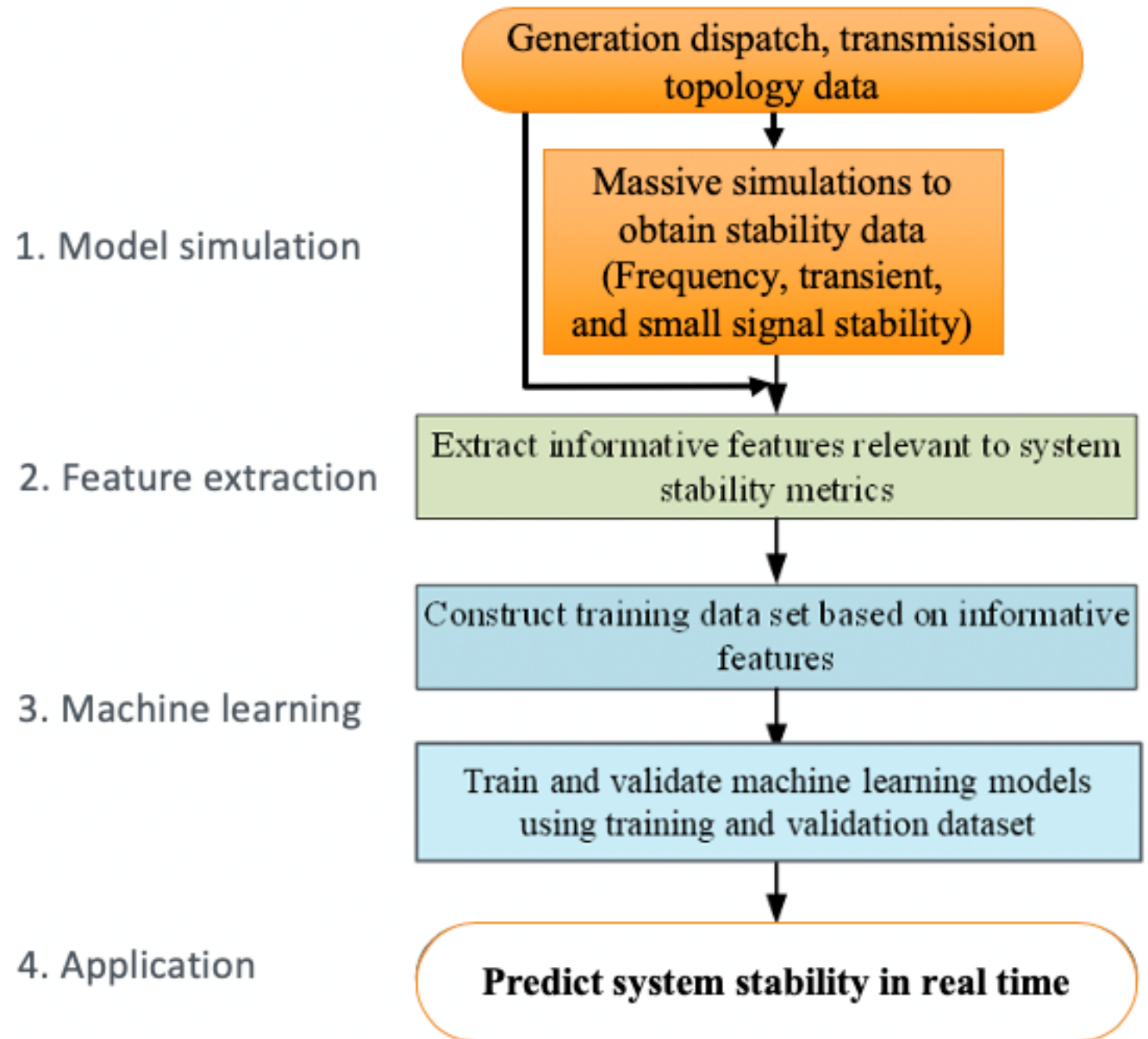

**Figure 18.** Framework of AI-based system stability prediction.

With AI, the assessment of power grid stability margins can be simplified [47, 53, 54]. AI can identify the nonlinear correlation between system stability margins and the system dispatch pattern, and then build a robust and accurate mapping between power flow data to stability margin indices [55]. Table 4 shows the inputs and outputs for stability prediction based on AI. These inputs are selected based on power system engineering principles and statistical results. AI algorithms to predict system stability can be commonly-used neural networks and random forests. Figure 18 shows the framework of the AI-based stability margin prediction, which follows a sequential process including feature extraction, model training, validation, and testing. The accurate stability margin in the training and testing dataset is obtained from the detailed time-domain simulation using dynamic models.

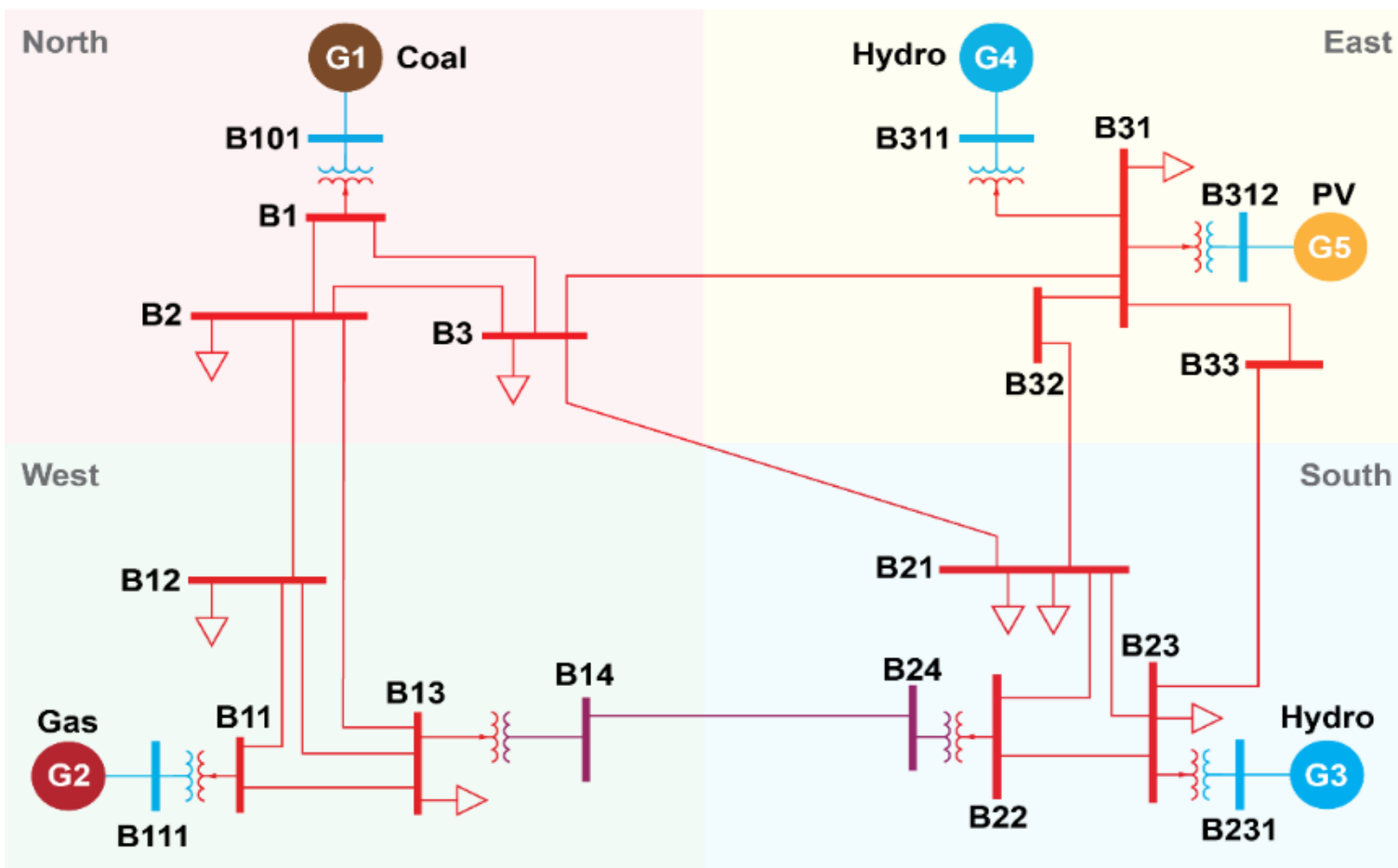

**Figure 19.** The 18-bus test system for AI-based stability prediction.

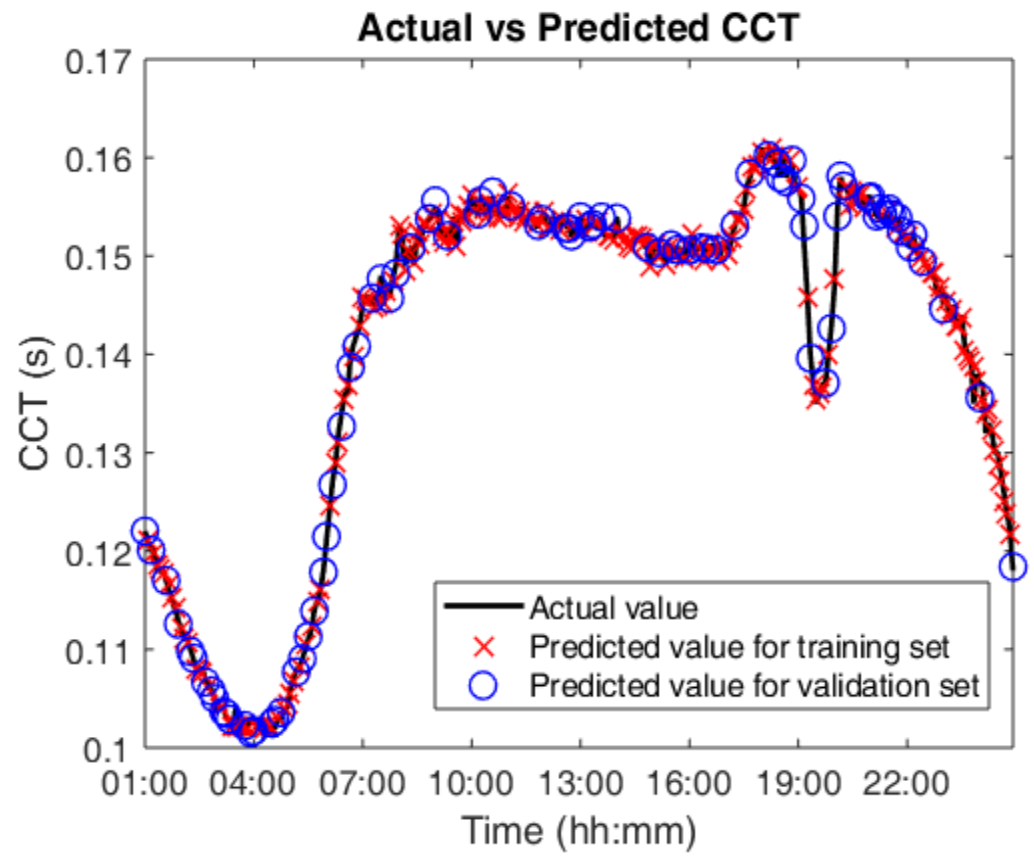

**Figure 20.** AI-based transient stability assessment results.

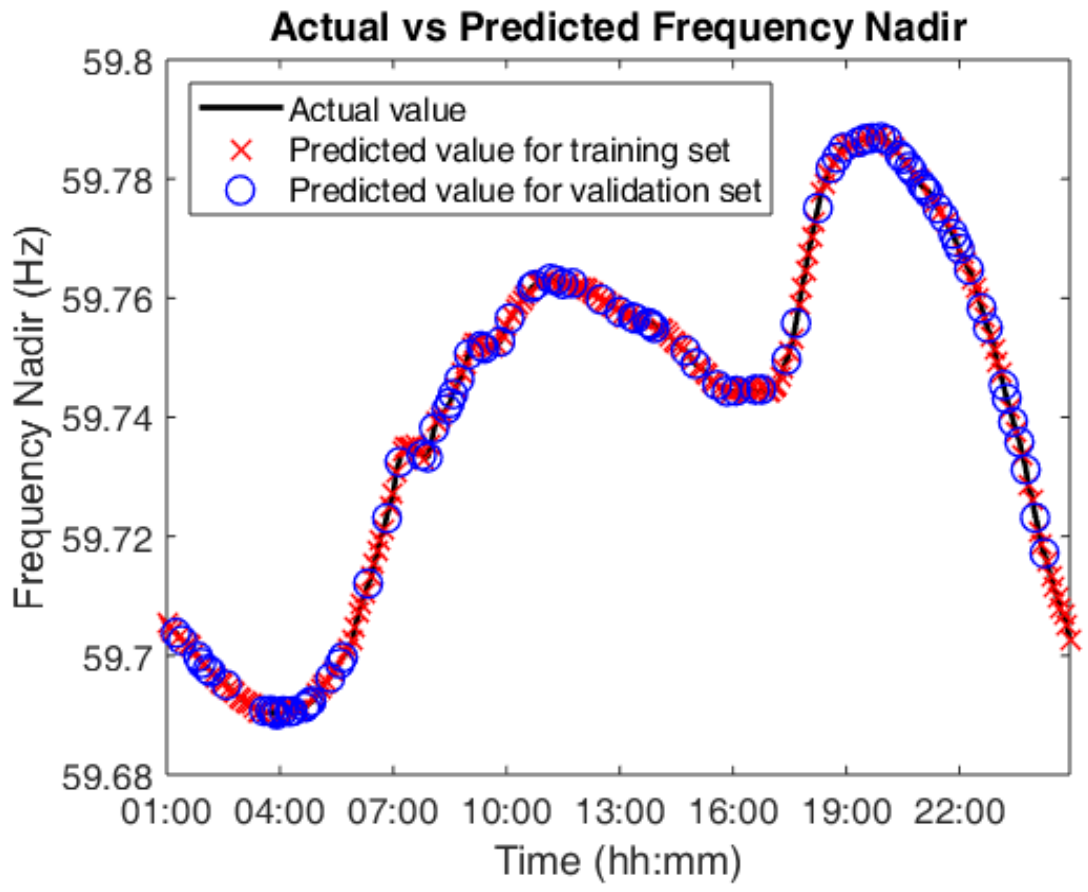

**Figure 21.** AI-based frequency stability assessment result (inertia change).

The AI-based stability assessment tool is tested on an 18-bus system shown in Figure 19. It includes four conventional synchronous generators and one PV power plant. It has 288 power flow scenarios (one scenario every 5 minutes for 1 day). 70 % of these scenarios, or 202 scenarios, are used in training, and the rest 30%, or 86 scenarios, are used for testing.

The transient stability assessment results are shown in Figure 20. It can be seen that AI can accurately predict transient stability quantified by critical clearing time (CCT). The frequency stability assessment result is shown in Figure 21 and Figure 22. In Figure 21, AI is used to predict the frequency nadir using system inertia as inputs. In this case, it is assumed that only system inertia changes and governors' statuses remain unchanged. In Figure 22, AI is used to predict the frequency nadir using both the inertia and the governors' statuses. The results show that the AI can consider both inertia and governors' statuses in predicting the system frequency nadir. Figure 23 and Figure 24 show the AI-based small-signal stability prediction performance. The result shows that AI can predict both the damping ratio and the oscillation frequency accurately.

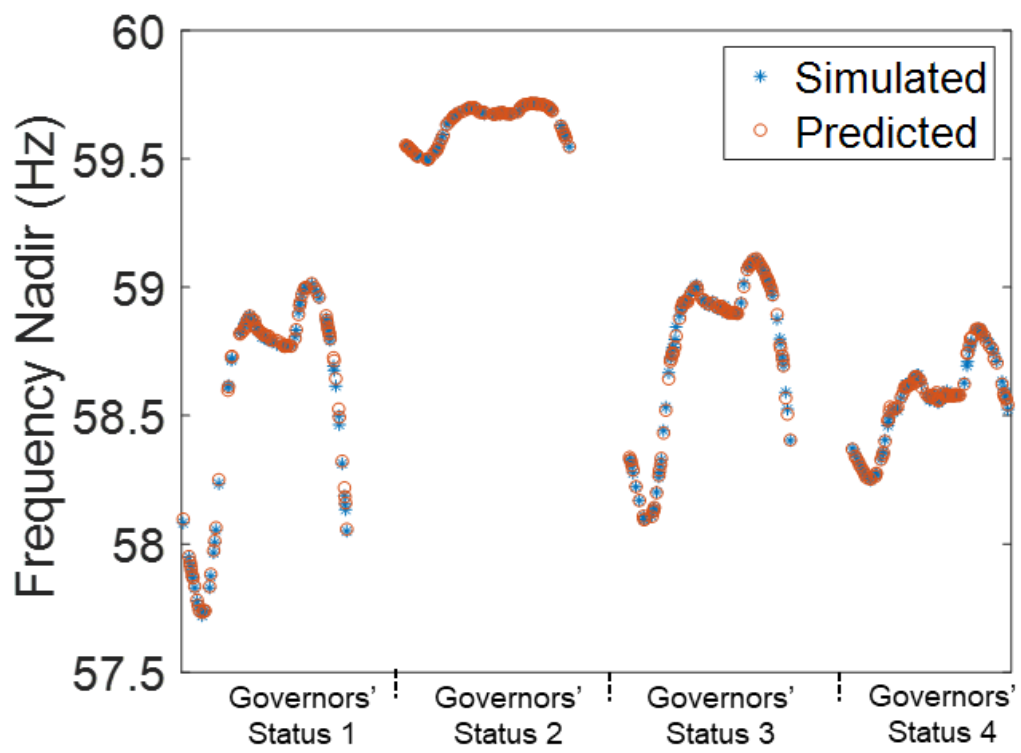

**Figure 22.** AI-based frequency stability assessment result (inertia and governor status change).

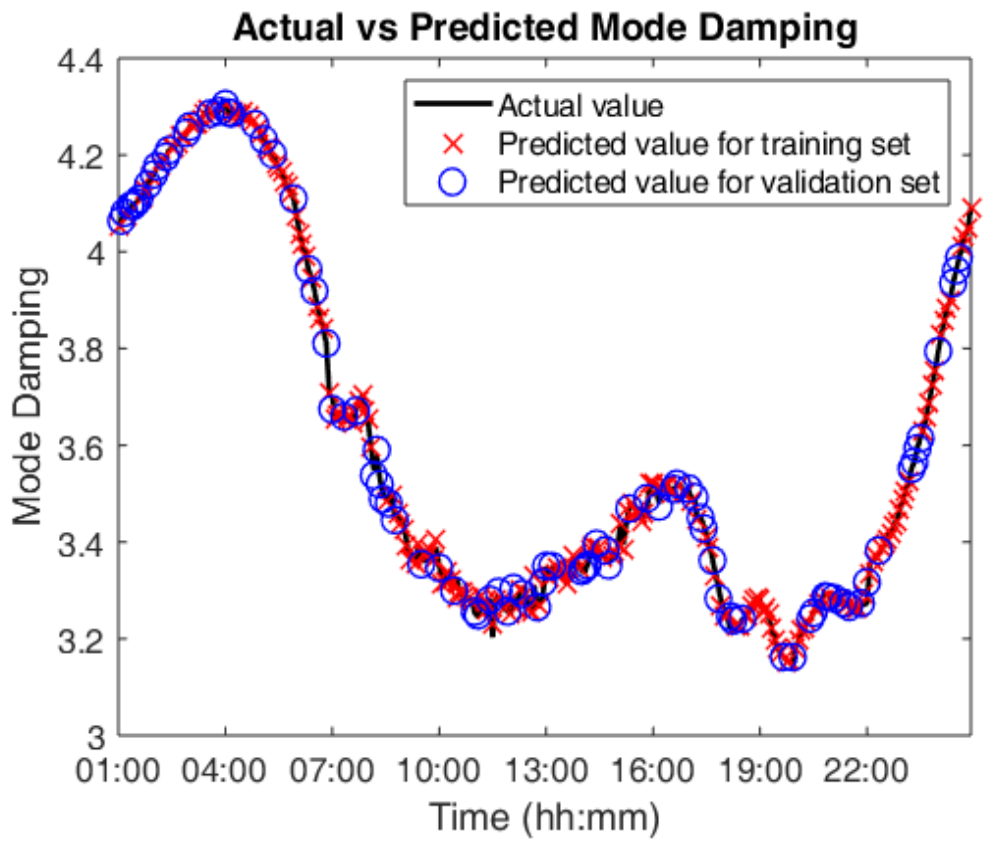

**Figure 23.** AI-based small signal stability assessment result (damping ratio).

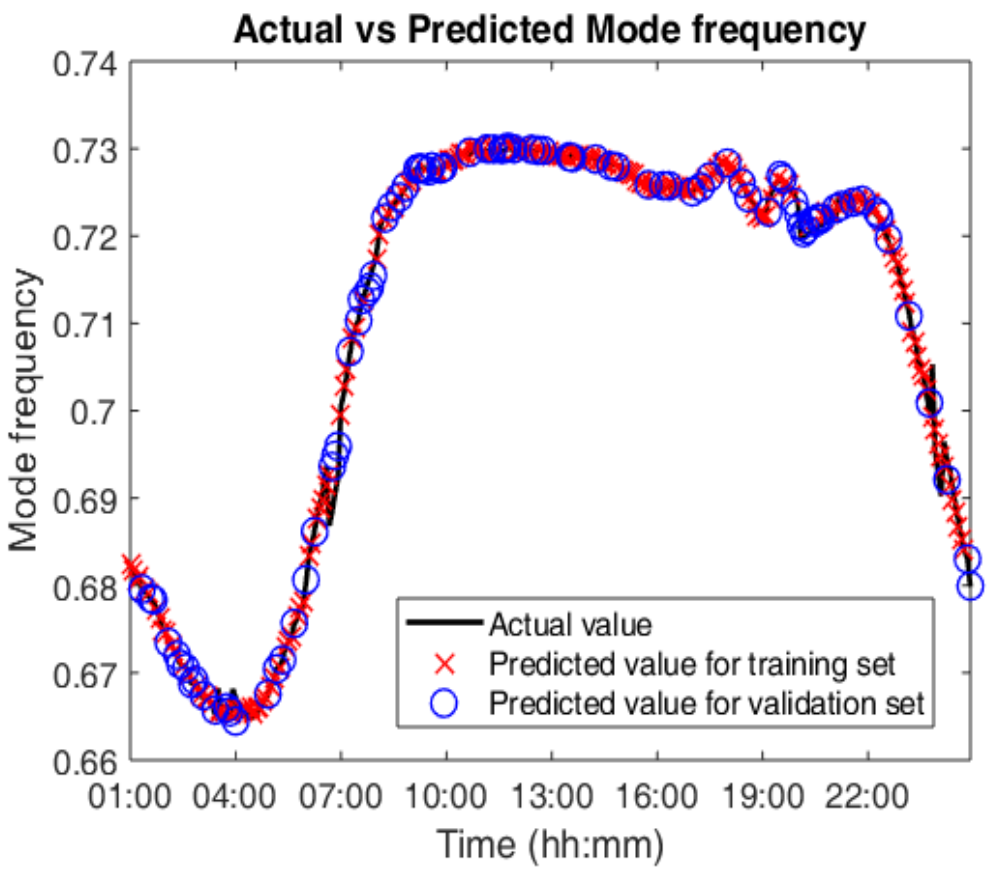

**Figure 24.** AI-based small signal stability assessment result (oscillation frequency).

Table 5 summarizes the accuracy of different AI-based stability assessment methods. It shows that in general, neural network has higher accuracy compared with random forests. One exception is in small signal stability assessment, in which random forests obtained slightly higher accuracy. To evaluate computation time savings, Table 6 compares the computation time of conventional model-based time-domain simulation and the AI-based methods. It can be seen that AI can significantly save computation time. This fast evaluation feature is especially useful for real-time stability assessment since the dynamic model may be unavailable or detailed simulation is too time-consuming.

**Table 5.** Accuracy of different testing of AI-based stability assessment.

| Stability | Metric | Estimation accuracy | |
|---|---|---|---|
| | | Random forests | Neural network |
| Frequency | Nadir | 98.30% | 99.72% |
| Transient | CCT | 98.44% | 99.29% |
| Small-Signal | Damping ratio | 98.61% | 98.59% |

**Table 6.** Comparison of simulation time using simulation and AI.

| Stabilities | Time for stability assessment (86 scenarios) | |
|---|---|---|
| | Simulation | AI-based |
| Transient stability | ~16 h | ~0.18 ms (with trained model) |
| Frequency stability | ~1 h | |
| Small signal stability | ~1 h | |

## 8. Conclusions

This paper presented a series of AI-based power grid applications developed on FNET/GridEye. These applications cover a wide range of areas that help improve power grid reliability, including event identification, event location and magnitude estimation, inertia estimation, data authentication, frequency control, and stability assessment. These AI applications have better performance compared with conventional approaches developed based on human experience, and have a high accuracy comparable to model simulation, indicating huge potentials in improving power grids situational awareness and stability. The primary reason of this improved performance is AI's capability in learning the non-linear relation between power grid measurements and stabilities indices from power grid big data. Further advancement of AI-based applications in power grids will rely on measurement data, accurate system modeling, next-generation AI and data analytics technologies. Future work will be continuing developing AI based applications to leverage more data from new sensors, next-generation AI technologies, to meet new demands and facilitate new paradigms for smarter grids.

**Author Contributions:** Y.L. (corresponding author) is the lead of all efforts in this paper. S.Y. wrote the original draft and designed the study in Section 6 and Section 7. W.W. conducted the study in Section 2. Y.C. performed the study in Section 3 and 4. S.L. performed the study in Section 5. H.L. and Y.S. did the study in Section 6. Y.Z., M.M., H.X., S.F., and L.Z. are contributors of the study in Section 7. K.S., C.C., H.Y., W.Y., W.Y., C.Z., Y.W., and J.D. contributed to the data collection for the study in Section 2, 3, 4, and 5. H.Y., H.J., and J.T. are the project lead of the study in Section 7. S.L. and K.S. reviewed and edited the manuscript.

**Funding:** This work was supported in part by the U.S. Department of Energy Solar Energy Technologies Office under Award # 34231 and 34224. Funding for this research was also provided by the NSF Cyber-Physical Systems (CPS) Program under award number 1931975. This work was also supported in part by the Engineering Research Center Program of the National Science Foundation and the Department of Energy under NSF Award Number EEC-1041877 and the CURENT Industry Partnership Program.

**Conflicts of Interest:** The authors declare no conflict of interest.